\documentclass[aps,prl,twocolumn,superscriptaddress,a4paper,nopacs]{revtex4}

\usepackage{graphicx}
\usepackage{amssymb}
\usepackage{amsmath}

\usepackage[colorlinks=true,citecolor=blue,linkcolor=magenta]{hyperref}
\hypersetup{
pdfauthor = {P. Maletinsky},
colorlinks = true, linkcolor = blue, urlcolor=blue, bookmarksnumbered =  true}

\newcommand{\ket}[1]{\left|#1\right>}

\usepackage{graphicx}

\begin{document}

\title{Quantitative nanoscale vortex-imaging using a cryogenic quantum magnetometer}

\author{L.~Thiel}
\affiliation{Department of Physics, University of Basel, Klingelbergstrasse 82, Basel CH-4056, Switzerland}
\author{D.~Rohner}
\affiliation{Department of Physics, University of Basel, Klingelbergstrasse 82, Basel CH-4056, Switzerland}
\author{M.~Ganzhorn}
\affiliation{Department of Physics, University of Basel, Klingelbergstrasse 82, Basel CH-4056, Switzerland}
\author{P.Appel}
\affiliation{Department of Physics, University of Basel, Klingelbergstrasse 82, Basel CH-4056, Switzerland}
\author{E.~Neu}
\affiliation{Department of Physics, University of Basel, Klingelbergstrasse 82, Basel CH-4056, Switzerland}
\author{B.~M\"uller}
\affiliation{Physikalisches Institut and Center for Quantum Science (CQ) in LISA$^+$, Universit\"at T\"ubingen, Auf der Morgenstelle 14, D-72076 T\"ubingen , Germany}
\author{R.~Kleiner} 
\affiliation{Physikalisches Institut and Center for Quantum Science (CQ) in LISA$^+$, Universit\"at T\"ubingen, Auf der Morgenstelle 14, D-72076 T\"ubingen , Germany}
\author{D.~Koelle}
\affiliation{Physikalisches Institut and Center for Quantum Science (CQ) in LISA$^+$, Universit\"at T\"ubingen, Auf der Morgenstelle 14, D-72076 T\"ubingen , Germany}
\author{P.~Maletinsky}
\email{patrick.maletinsky@unibas.ch}
\affiliation{Department of Physics, University of Basel, Klingelbergstrasse 82, Basel CH-4056, Switzerland}

\date{\today}

\begin{abstract}
Microscopic studies of superconductors and their vortices play a pivotal role in our understanding of the mechanisms underlying superconductivity\,\cite{Essmann1967,Mannhart1987,Bending1999,Hess1989, Hoffman2002, Roditchev2015,Fischer2007}. 
Local measurements of penetration depths\,\cite{Luan2010} or magnetic stray-fields\,\cite{Kirtley2012} enable access to fundamental aspects of superconductors such as nanoscale variations of superfluid densities\,\cite{Luan2010} or the symmetry of their order parameter\,\cite{Hardy1993}.
However, experimental tools, which offer quantitative, nanoscale magnetometry and operate over the large range of temperature and magnetic fields relevant to address many outstanding questions in superconductivity, are still missing.
Here, we demonstrate quantitative, nanoscale magnetic imaging of Pearl vortices in the cuprate superconductor YBa$_2$Cu$_3$O$_{7-\delta}$, using a scanning quantum sensor in form of a single Nitrogen-Vacancy (NV) electronic spin in diamond\,\cite{Rondin2012,Tetienne2014b,Maletinsky2012}.
The sensor-to-sample distance of $\approx10~$nm we achieve allows us to observe striking deviations from the prevalent monopole approximation\,\cite{Auslaender2008} in our vortex stray-field images, while we find excellent quantitative agreement with Pearl's analytic model\,\cite{Pearl1964}. 
Our experiments yield a non-invasive and unambiguous determination of the system's local London penetration depth, and are readily extended to higher temperatures and magnetic fields. 
These results demonstrate the potential of quantitative quantum sensors in benchmarking microscopic models of complex electronic systems and open the door for further exploration of strongly correlated electron physics using scanning NV magnetometry.
\end{abstract}

\maketitle

\begin{figure}[h!]
\includegraphics[width=8.6cm]{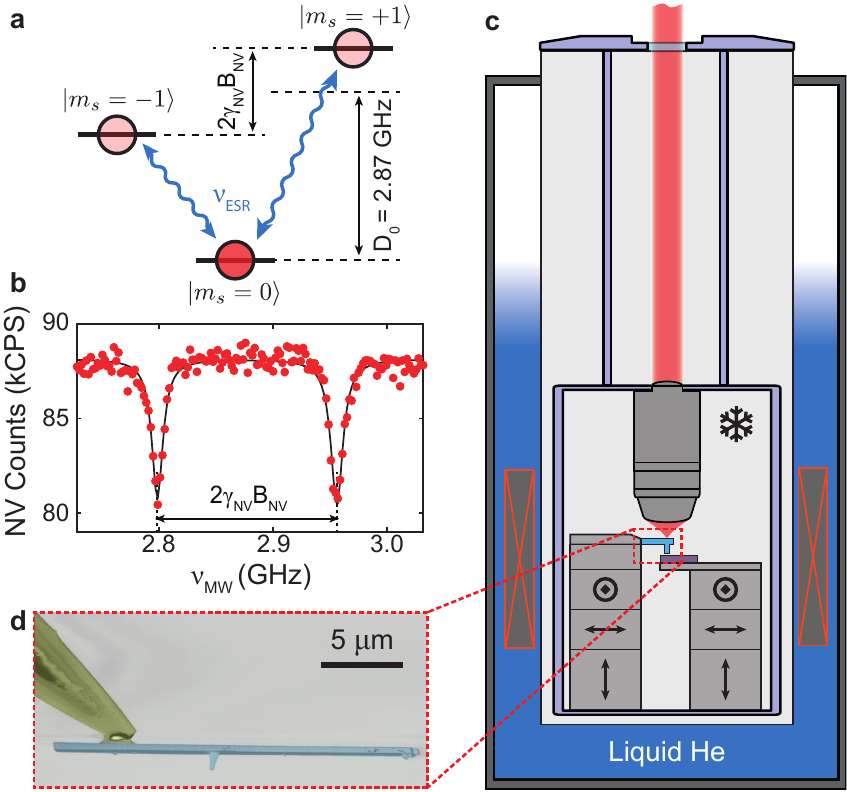}
\caption{\label{FigOverview} {\bf Basis of NV magnetometry and overview of experimental apparatus}. 
{\bf a}~Ground state spin levels of the optically active, negatively-charged diamond NV centre, which exhibit spin-dependent fluorescence rates (red circles) and optical spin pumping under green excitation (see text). Microwave magnetic fields of frequency $\nu_{\rm ESR}$ can drive electron spin resonances (ESR), which are optically detectable. 
{\bf b}~Typical NV ESR trace obtained from a single NV in a diamond scanning probe at $4.2~$K. Fluorescence count-rate, ESR contrast and linewidth yield a magnetic field sensitivity of $11.9~\mu$T$/\sqrt{\rm Hz}$.
{\bf c}~Layout of the cryogenic, scanning NV magnetometer. Tip and sample scanning are enabled by three-axis coarse and fine positioning units and NV fluorescence is collected through a taylor-made low-temperature objective. The microscope resides in a liquid $^4$He bath at temperature $4.2~$K.
{\bf d} False-color electron microscope image of an all-diamond scanning probe as used here. The NV sensor spin is located at the apex of the nanopillar visible in the center of the diamond cantilever.
}
\end{figure} 

Understanding the microscopic mechanisms of superconductivity is a central topic of modern condensed matter physics and vital for the development of novel superconducting materials of technological and scientific relevance.
A multitude of approaches have been developed in the past to study superconductivity on the micro- and nano-scale including magnetic decoration\,\cite{Essmann1967}, electron microscopy\,\cite{Mannhart1987} 
or magneto-optical imaging\,\cite{Bending1999}. 
The quest to access physical quantities such as stray-fields or the superfluid density at the nanoscale has driven developments of scanning probe technologies, where 
seminal results have been achieved by approaches such as scanning tunnelling microscopy\,\cite{Hess1989, Hoffman2002, Roditchev2015,Fischer2007}
scanning Hall-probes\,\cite{Dinner2005}, scanning superconducting quantum interference devices\,\cite{Kalisky2010,Vasyukov2013} or magnetic force microscopy\,\cite{Auslaender2008}. While such approaches continue to deliver unprecedented insight into the physics of superconductors and condensed matter systems in general, each also suffers from  drawbacks such as limited temperature and magnetic field range, limited spatial resolution, high invasiveness, or the inability to address electrically isolating samples. A promising new approach to nanoscale magnetic field imaging was recently opened by the advent of single-spin based nanoscale magnetometers,
in particular using single electronic spins in diamond\,\cite{Taylor2008, Balasubramanian2008,Rondin2014}. 
Such sensors have the potential to overcome the above-mentioned limitations and promise to form a highly valuable addition to the existing toolset for nanoscale studies of complex electronic systems\,\cite{Bouchard2009}. 
However, cryogenic operation -- a vital requirement for application to many open problems in condensed matter physics -- to date has not been demonstrated for this technology. 

Magnetometry with NV centre spins in diamond builds on the fact that the negatively charged NV crystal defect is optically active and possesses a spin-1 ground state, with a highly spin-dependent fluorescence rate (Fig.\,\ref{FigOverview}a). 
Optical excitation at $532~$nm pumps the NV into the $\ket{m_s=0}$ state, from which resonant microwave magnetic fields can promote it to one of the less-fluorescent $\ket{m_s=\pm1}$ states. In combination, these properties allow for efficient, optical detection of NV electron spin resonance (ESR)\,\cite{Rondin2014}, as shown in Fig.\,\ref{FigOverview}b for a single NV in our cryogenic apparatus. NV magnetometry is based on the magnetic dispersion of the $\ket{m_s=\pm1}$ states, which leads to a linear splitting of $2\gamma_{\rm NV}B_{\rm NV}$ of the two NV ESR dips, where $\gamma_{\rm NV}=28~$MHz/mT and $B_{\rm NV}$ is the magnetic field along the NV symmetry axis. Magnetometry based on such direct ESR measurements yields typical sensitivities of few $\mu$T$/\sqrt{\rm Hz}$\,\cite{Rondin2014} --- largely sufficient for the magnetic targets we address in this work. 
For weaker magnetic sources, sensitivity can be further enhanced by coherent spin manipulation and dynamical decoupling\,\cite{Taylor2008}, which for the type of devices used here improves sensitivity to the nT$/\sqrt{\rm Hz}$-range\,\cite{Maletinsky2012}.

Our experimental setup (Fig.\,\ref{FigOverview}c) consists of a combined confocal and atomic force microscope (AFM) operating in a 
$^4$He bath cryostat with a base-temperature of $4.2~$K (see Methods). This approach yields high cooling-power, minimal vibrations for AFM operation and allows for full, vectorial magnetic field control up to $0.5~$T.
The microscope is placed in a $^4$He buffer-gas filled housing, which is directly immersed in the liquid $^4$He bath for  cooling. Optical access is provided by a quartz window on top of the cryostat and a cryogenic objective, rigidly attached to the microscope head. The single NV spin for magnetometry is embedded in an individual, all-diamond scanning probe within $\sim10~$nm from the tip\,\cite{Maletinsky2012} (Fig.\,\ref{FigOverview}d). This approach ensures high NV photon collection efficiencies for sensitive magnetometry and close proximity of the NV to the sample for high imaging resolution\,\cite{Maletinsky2012,Appel2015}.

\begin{figure}[t!]
\includegraphics[width=8.6cm]{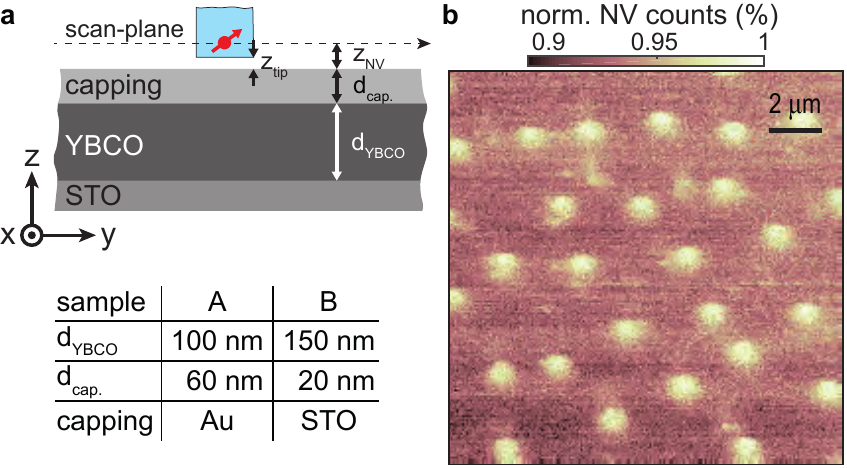}
\caption{\label{FigVorticesIsoB} {\bf Ensemble vortex imaging and sample design.} 
{\bf a}~Layout of sample and scanning NV sensor. 
The superconducting YBCO film of thickness $d_{\rm YBCO}$ was grown on an STO substrate (see text and Methods) and covered by a protective layer of thickness $d_{\rm cap.}$. Key parameters for samples A and B are summarised in the table inset.
The red arrow and blue structure indicates the NV spin and diamond nanopillar at distances $z_{\rm NV}$ and $z_{\rm tip}$ from the sample surface, respectively. 
{\bf b }~Iso-magnetic field image of an ensemble of vortices in sample A imaged at $B=0$ after field-cooling in $B_z^{\rm f.c.}=0.4~$mT. 
The microwave driving frequency 
$\nu_{\rm MW}$ was fixed to the zero field NV ESR at $2.87~$GHz and NV fluorescence monitored while scanning the sample. Bright areas indicate regions where the NV Zeeman shift exceeds the ESR half linewidth of $6~$MHz, i.e. where 
$B_{\rm NV}>0.22~$mT.
}
\end{figure}

We performed our experimental demonstration of cryogenic, nanoscale NV magnetometry by imaging Pearl\hphantom{} vortices in the prototypical superconductor YBa$_2$Cu$_3$O$_{7-\delta}$ (YBCO) with a transition temperature $T_c\approx 89\,$K. The type II superconductor YBCO is amongst the best-studied high-T$_c$ superconductors, forms a model system for strongly correlated electron physics, and hosts vortices which generate non-trivial stray magnetic-fields\,\cite{Auslaender2008}. 
It therefore offers an ideal testbed to demonstrate the applicability of our method 
and allows us to benchmark existing models for vortex stray-fields against each other. 
The samples we studied (Fig.\,\ref{FigVorticesIsoB}a) consist of a thin, single-crystalline film of YBCO with thickness $d_{\rm YBCO}$, which was grown epitaxially 
on a SrTiO$_3$ (STO) substrate (see Methods). To prevent degradation of YBCO's superconducting properties the films are covered with a protective capping-layer of thickness $d_{\rm cap.}$. We studied two samples (denoted ``A'' and ``B'') with $d_{\rm YBCO}=100~$nm ($150~$nm) and $d_{\rm cap.}=60~$nm ($20~$nm), respectively (see table in Fig.\,\ref{FigVorticesIsoB}a). The samples were mounted 
nearby a stripline for microwave (MW) delivery 
for NV spin manipulation
and a heater for temperature control. 
A gold bonding wire was placed within few microns of the sample and connected to the MW leads for ESR driving. 
To nucleate vortices, 
 we field-cooled the samples from a temperature $T>T_c$ to the system base temperature in an external magnetic field $B_z^{\rm f.c.}=0.4~$mT, which we applied normal to the sample surface.

\begin{figure*}[t!]
\includegraphics[width=17cm]{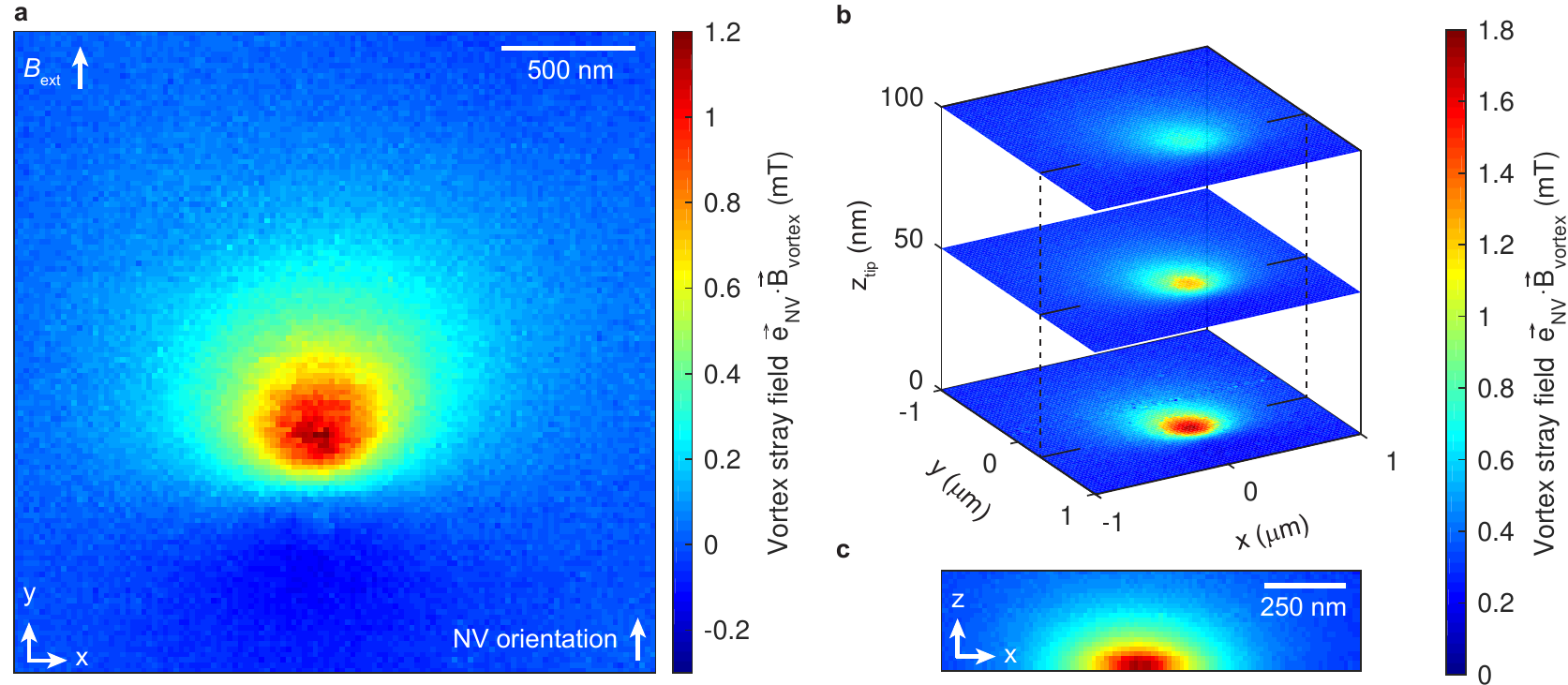}
\caption{\label{FigVortexFullField} {\bf Quantitative mapping of single-vortex stray magnetic fields.}
{\bf a}~Image of the magnetic stray-field from a single vortex in sample A obtained with the NV magnetometer in AFM contact. The stray-field projection onto the NV axis was obtained by measuring the Zeeman-splitting in optically detected ESR (Fig.\,\ref{FigOverview}b) at each of the $120\times120$ pixels of the scan. 
The width of the observed vortex stray-field is set by $\Lambda$ (the Pearl length), which is much bigger than our estimated spatial resolution of $10~$nm (see text).
{\bf b}~Three-dimensional reconstruction of $B_{\rm NV}$ obtained as in {\bf a} with a different diamond tip. Scans were performed at two out-of-contact heights of $50~$nm and $100~$nm, as indicated.
{\bf c}~Vertical scan through the vortex stray-field in the $x/z$-plane indicated in {\bf b}. 
}
\end{figure*}

To image the resulting vortex distribution, we first performed a large-area iso-magnetic field image using our scanning NV magnetometer. To that end, we fixed the MW driving  frequency $\nu_{\rm MW}$ to the zero-field NV-ESR frequency of $\nu_{\rm ESR}=2.87~$GHz and scanned the sample below the stationary NV, whose fluorescence we constantly interrogated. Whenever a vortex was scanned below the NV, the vortex stray magnetic field shifted $\nu_{\rm ESR}$ away from $\nu_{\rm MW}$, resulting in an increased NV fluorescence rate. Bright spots in Fig.\,\ref{FigVorticesIsoB}b therefore signal the presence of individual vortices in the sample. 
Given our cooling-field of $B_z^{\rm f.c.}=0.4~$mT and the magnetic flux-quantum $\Phi_0=h/2e=2.07~$mT$\mu$m$^2$ (with Planck's constant $h$ and the electron charge $e$), one expects a vortex-density $B_z^{\rm f.c.}/\Phi_0=0.19\mu$m$^{-2}$, i.e. $43~$vortices in our scan-range of $15~\mu$m$\times 15~\mu$m in fair agreement with the $27$~vortices observed in Fig.\,\ref{FigVorticesIsoB}.

Further insight into individual vortices can be gained by full, quantitative mapping of the magnetic stray field emerging from a single vortex, which we performed on sample A. 
To that end, we focussed on a spatially isolated vortex and conducted a scan with our NV magnetometer, where we obtained $B_{\rm NV}$ by measuring the Zeeman-splitting in optically detected ESR (Fig.\,\ref{FigOverview}b) at each pixel of the scan. 
The vortex was nucleated as before and imaging was performed in a weak bias-field $B_{\rm NV, bias}=0.45~$mT, which we applied to determine the sign of the measured fields. The resulting image (Fig.\,\ref{FigVortexFullField}a) yields a map of the projection of the vortex stray-field $\vec{B}_{\rm vortex}(x,y,z)$ onto the NV spin-quantisation axis, $\vec{e}_{\rm NV}=(0,\sqrt{2},1)/\sqrt{3}$ (which we independently determined prior to the scan, see SOM).
The non-zero angle between $\vec{e}_{\rm NV}$ and the sample normal $\vec{e}_z$ leads to an asymmetry of the observed stray-field image, which would otherwise be rotationally symmetric in the $x/y$ plane.
(Note that due to strong twinning in our thin film sample, the in-plane London penetration depth $\lambda_L$ is essentially isotropic). 
Our quantitative image allows us to directly determine the total magnetic flux $\Phi_{\rm B, meas.}\approx0.79~$mT$\mu$m$^2$ 
enclosed by
the vortex 
by spatial integration of the data in Fig.\,\ref{FigVortexFullField}a.
This value is in good agreement with the expected flux 
after taking into account the NV orientation and the finite integration area of our measurement (see SOM).

The outstanding stability of our cryogenic NV magnetometer further allows us to perform a full, three-dimensional mapping of $\vec{e}_{\rm NV}\cdot\vec{B}_{\rm vortex}(x,y,z)$. To that end, 
we released AFM feedback and scanned the sample at well-defined distances of $z_{\rm tip}=50~$nm and $100~$nm from the diamond-tip. The resulting slices of $\vec{B}_{\rm vortex}(x,y,z=z_{\rm NV})$ (Fig.\,\ref{FigVortexFullField}b), together with an independently measured map of $\vec{B}_{\rm vortex}(x,y,z)$ along an $x/z-$plane (Fig.\,\ref{FigVortexFullField}c) provide complete, quantitative information about the stray magnetic field emerging from the vortex. 
Importantly, our data shows variations of $\vec{B}_{\rm vortex}(x,y,z)$ with $z$ down to the smallest values of $z$ (where $z_{\rm tip}\approx0$), which provides evidence that our imaging is not limited by detector size.
In fact, our magnetometer can detect changes in $\vec{B}_{\rm vortex}(x,y,z)$ with a resolution which is determined by the spatial extent of the NV's electronic wave-function, i.e. on lengthscales $<1~$nm. 
As a result, our imaging resolution for magnetic texture on the sample surface is solely set by 
$z_{\rm NV}$ (the distance between NV and sample surface, see. Fig.\,\ref{FigVorticesIsoB}a)
which in AFM contact is $\sim10~$nm\,\cite{Appel2015} (see below).

\begin{figure}[t!]
\includegraphics[width=8.6cm]{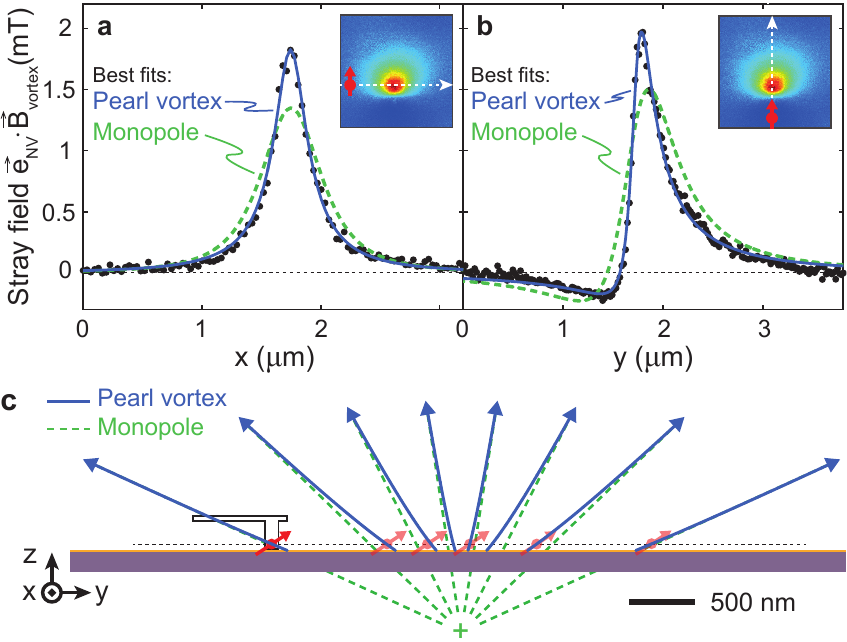}
\caption{\label{FigVortexAnalysis} {\bf Quantitative stray-field analysis and determination of the London penetration depth.}
{\bf a} and {\bf b}~High-resolution measurements of the vortex stray-field ($\vec{e}_{\rm NV}\cdot\vec{B}_{\rm vortex}$) 
recorded in sample B along the horizontal (``$x$'') and vertical (``$y$'') symmetry-axes of a single vortex, as illustrated in the insets. Blue (green dashed) lines represent best fits to a Pearl vortex (magnetic monopole), respectively. 
From the Pearl-vortex fit, we determine 
$\lambda_L=251~\pm14$nm 
the bulk London penetration depth which can not be obtained unambiguously using the monopole model. 
{\bf c}~Vortex magnetic field lines close to the superconductor surface within the monopole (green dashed) and Pearl (blue) approximations, illustrating the strong discrepancy between the two models close to the surface ($z_{\rm NV}\ll\lambda_L$). In all panels, the red arrow illustrates the orientation of the NV with respect to the sample. 
}
\end{figure}

A central distinguishing feature of NV magnetometry is it's ability to provide quantitative measures of magnetic fields on the nanoscale. 
Here, this feature allows us to test and discriminate existing models for vortex stray fields and to determine local properties of our superconducting sample.
To that end, we conducted separate, high resolution line-scans of $\vec{B}_{\rm vortex}(x,y,z)$ along the symmetry axes of an individual vortex (nucleated by field-cooling in $B_{z}^{\rm f.c.}=0.2~$mT) in sample B. 
The resulting measurements of $\vec{B}_{\rm vortex}(x)$ and $\vec{B}_{\rm vortex}(y)$ (Fig.\,\ref{FigVortexAnalysis}a and b, respectively) form the basis for our subsequent, quantitative analysis.

A widely used model for the stray magnetic field of a superconducting vortex consists of the field of a virtual magnetic monopole of strength $2\Phi_0$, located within a distance $\lambda_L$ from the surface inside the superconductor\,\cite{Pearl1964,Auslaender2008,Carneiro2000}. 
This monopole approximation is simple and often appropriate, but also suffers from strong limitations. It does not allow for an independent determination of $\lambda_L$ and $z_{\rm NV}$ (i.e. changes in $\lambda_L$ cannot be distinguished from changes in $z_{\rm NV}$) and it breaks down at small distances$~\ll\lambda_L$ from the superconductor. 
Indeed, in our attempts to fit the data (keeping $\vec{e}_{\rm NV}$ and $\Phi_0$ fixed and varying the vortex position), the monopole failed to yield a satisfactory fit (Fig.\,\ref{FigVortexAnalysis}a\&b, green dashed lines). This discrepancy is a consequence of the close proximity of the NV to the sample ($z_{\rm NV}+d_{\rm cap.}\ll\lambda_L$) and an opportunity to test the validity of more refined models for vortex stray fields. 

A more accurate description of the vortex stray-field is offered by Pearl's approach\,\cite{Pearl1964,Carneiro2000} to 
obtain the distribution of superconducting currents (and thereby the stray magnetic field) around vortices in thin-film superconductors. 
The resulting fit of this Pearl-vortex stray-field to our data (Fig.\,\ref{FigVortexAnalysis}a\&b, blue lines) indeed shows excellent quantitative agreement. 
Importantly, and in contrast to the monopole approximation, this fit, paired with the high signal-to-noise ratio of our data, allows us to independently determine $\lambda_L$ and $h_{\rm NV}$, the vertical distance between the NV and the Pearl vortex at the center of the YBCO film. 
From the fit (see SOM), we find a Pearl length 
$\Lambda=2\lambda_L^2/d_{\rm YBCO}=840\pm20~$nm 
and $h_{\rm NV}=104\pm2~$nm. Here,  $h_{\rm NV}$ is related to the net NV-to-sample standoff distance $z_{\rm NV}=h_{\rm NV}-d_{\rm YBCO}/2-d_{\rm cap.}=9\pm3.5~$nm 
 --- our ultimate imaging resolution for magnetism on the sample surface. 
Our measurement of $\Lambda$ yields a bulk penetration depth $\lambda_L=251\pm14~$nm, 
 which is consistent with previously reported values\,\cite{Wolbing2014,Auslaender2008} and provides proof for the validity of our model and method. 

The analysis of our experimental data is exemplary for the great potential the quantitative aspects of NV magnetometry hold for future applications in studying complex condensed matter systems. 
Our quantitative fits allowed us to locally determine the absolute value of $\lambda_L$ ---  a quantity which is notoriously hard to measure\,\cite{Luan2010} but of high interest due to its direct link to 
structure of the superconducting gap\,\cite{Hardy1993}.  
Furthermore, our analysis allowed us to draw a clear distinction between two alternative models for vortex stray-fields 
--- in analogy, such analysis should in the future allow us to discriminate between competing models for magnetic order in a variety of condensed matter systems\,\cite{Young2012,Xia2006}.
Non-invasiveness of the probe will be a key requirement and warrants discussion of potential, unwanted heating effects due to NV laser excitation and microwave driving. For YBCO, we have repeated vortex imaging for increasing laser powers where even for the highest values $\approx2~$mW (see SOM), 
were able to image vortices without observing signs of ``vortex-dragging''\,\cite{Auslaender2008}.
For even more fragile samples, there is ample margin to further reduce potential heating effects by employing resonant NV laser excitation (requiring nW power-levels\,\cite{Robledo2011}), all-optical spin manipulation\,\cite{Yale2013} to eliminate MWs, or pulsed ESR detection\,\cite{Dreau2011} with projected laser duty cycles $<1\%$ for the longest reported NV coherence times\,\cite{Balasubramanian2009}. 
The resulting, sub-nW average heating powers we project compare favourably to established approaches to studying strongly correlated electron systems\,\cite{Xia2006} at ultralow temperatures. 

In conclusion, we have demonstrated for the first time operation of an NV-based scanning probe magnetometer operating under cryogenic conditions. We combined nanoscale spatial resolution and quantitative, non-invasive magnetic imaging to map the stray magnetic fields of individual Pearl\hphantom{} vortices in superconducting thin films with high sensitivity and spatial resolution. Our results establish NV magnetometry as a powerful tool to address complex, electronic systems through nanoscale magnetic field imaging. Specifically, our NV magnetometer could in the future be used to address the elusive pseudogap phase\,\cite{Fischer2007,Timusk1999} 
of high-T$_c$ superconductors or inhomogeneous superconductors at the nanoscale\,\cite{Kresin2006}. Both problems are largely inaccessible to present day nanoscale sensing technologies due to a lack of spatial resolution or sensitivity or the ability to operate at elevated temperatures. Further extending our measurement capabilities through spin-relaxometry\,\cite{Kolkowitz2015} or quantum-sensing based on dynamical decoupling\,\cite{DeLange2011} is another exciting avenue that would drastically enhance the dynamic range of our sensor and therefore yield access to the nontrivial dynamical properties of individual vortices\,\cite{Embon2015}. 
The resulting projected dynamical range and sensitivity, together with the spatial resolution and quantitative aspects demonstrated here will open the door for studying unexplored aspects of quantum matter much beyond applications in superconductivity\,\cite{Levitov2015,Young2012} demonstrated here.

{\it Note added:} During completion of this work, we became aware of related results on cryogenic NV magnetometry\,\cite{Pelliccione2015}.

\section{Methods}

{\bf Sample fabrication:} 

We fabricated epitaxially grown $c$-axis oriented YBa$_2$Cu$_3$O$_{7-\delta}$ (YBCO) thin films on SrTiO$_3$ (STO) single crystal (001)-oriented substrates by pulsed laser deposition (PLD), followed either by in-situ electron-beam-evaporation of Au at room temperature (sample A) or by in-situ epitaxial growth of an STO cap layer by PLD (sample B).
For details on PLD growth of our YBCO films on STO substrates, and their structural and electric transport properties see Refs.~[\onlinecite{Werner10,Scharinger12}].
In brief, our YBCO films typically yield $0.1^\circ$ full width half maximum of the rocking curve at the (005) x-ray diffraction peak, have an inductively measured transition temperature $T_c=89\,$K and normal state resistivity $\rho\approx 50\,\mu\Omega$cm at $T=100\,$K.

The thicknesses 
of the samples were determined through in-situ high-pressure reflection high-energy electron diffraction (RHEED) 
for the YBCO and STO films. With the $c$-axis lattice parameters, as measured via x-ray diffraction, we then obtain the thicknesses for the YBCO and STO films quoted in the text. The Au growth rate was determined via a quartz-crystal monitor. 
For all layers, we estimated the error in our thickness determination to $2-3\,\%$.

{\bf Experimental setup:} 

Our experimental setup is based on a low-temperature, tuning-fork based AFM (Attocube, attoLIQUID~1000). The microscope is housed in a low-vibration, Nitrogen-free liquid $^4$He bath cryostat with a base temperature of $4.2~$K, which is equipped with a 3D vector magnet ($0.5~$T in all directions, Janis). Optical access to the AFM tip is provided by an achromatic, low temperature compatible objective (Attocube LT-APO/VISIR/0.82, $0.82~$NA), and a home-built confocal microscope, directly mounted on top of the cryostat.
The microscope head (including the sample, the diamond tip and the objective) is located in a housing filled with $^4$He buffer gas. The housing is directly immersed in the liquid $^4$He bath.
The diamond tip is attached to a tuning fork for force-feedback in AFM, which is provided by commercial electronics (Attocube, ASC500). Two separate positioning units (Attocube, ANSxyz50 on top of ANPxyz51) provide individual high accuracy positioning of both, sample and sensor. Temperature control of the sample above $4.2~$K is provided by a resistive heater (IMS, ND3-1206EW1000G) and a temperature contoller (LakeShore, Model$~355$).
Excitation light for NV fluorescence is provided by a solid-state laser at $532~$nm (LaserQuantum, GEM532). Red fluorescence photons are coupled into a single-mode fibre guided to an avalanching photo diode for counting (Excelitas, SPCM-ARQH-13). Data acquisition and scan control is achieved using a digital acquisition card (National Instruments, NI-6602 and NI-6733) and a Matlab-based experiment control software.
Microwave signals for spin manipulation are generated by a signal generator (Rohde\&Schwarz, SMB100A), amplified (Minicircuit, ZHL-42W+)  and delivered to the NV centre using a gold wire (diameter, $25~\mu$m) positioned accross the sample.

\section{Acknowledgements}
We thank V.~Jacques, A.~H\"ogele and S.D.~Huber for fruitful discussions and valuable feedback on the manuscript. We further acknowledge Attocube systems for excellent support and the joint development of the microscope system used here.  We gratefully acknowledge financial support from SNI; NCCR QSIT; SNF grants 143697 and 155845; and EU FP7 grant 611143 (DIADEMS). 

%
%

\bibliographystyle{apsrev4-1}
\bibliography{BibVortexImagingALL}

\newpage

\section*{Supplementary Information}

The following supplementary material is divided into five sections. Each section provides background information related to specific topics of the main text. The sections are not built upon each other and can be read independently. 

Section~S1 provides details of our experimental setup. In Sect.~S2, we discuss sample growth and the determination of layer-thicknesses in our samples.
In Sect.~S3 we describe our procedure to experimentally determine the magnetic flux through the vortex measured in Fig.3a of the main text. 
Section~S4 describes our procedure to determine NV orientation and align our external magnetic field to the NV if necessary.
Section~S5 presents AFM data on our sample which we used to assess sample roughness and vibrational stability of our setup.
Finally, Sect.~S5 describes our evaluation of the invasiveness of our method, which we conducted through vortex images at increasing laser powers.

\section{S1. Experimental setup and diamond tip}
\label{SectSetup}

Our experimental setup is based on a low-temperature, tuning-fork based AFM (Attocube, attoLIQUID~1000). The microscope is housed in a low-vibration, Nitrogen-free liquid $^4$He bath cryostat with a base temperature of $4.2~$K, which is equipped with a 3D vector magnet ($0.5~$T in all directions, Janis). Optical access to the AFM tip is provided by an achromatic, low temperature compatible objective (Attocube LT-APO/VISIR/0.82, $0.82~$NA), and a home-built confocal microscope, directly mounted on top of the cryostat.
The microscope head (including the sample, the diamond tip and the objective) is located in a housing filled with $^4$He buffer gas. The housing is directly immersed in the liquid $^4$He bath.
The diamond tip is attached to a tuning fork for force-feedback in AFM, which is provided by commercial electronics (Attocube, ASC500). Two separate positioning units (Attocube, ANSxyz50 on top of ANPxyz51) provide individual high accuracy positioning of both, sample and sensor. Temperature control of the sample above $4.2~$K is provided by a resistive heater (IMS, ND3-1206EW1000G) and a temperature contoller (LakeShore, Model$~355$).
Excitation light for NV fluorescence is provided by a solid-state laser at $532~$nm (LaserQuantum, GEM532). Red fluorescence photons are coupled into a single-mode fibre guided to an avalanching photo diode for counting (Excelitas, SPCM-ARQH-13). Data acquisition and scan control is achieved using a digital acquisition card (National Instruments, NI-6602 and NI-6733) and a Matlab-based experiment control software.
Microwave signals for spin manipulation are generated by a signal generator (Rohde\&Schwarz, SMB100A), amplified (Minicircuit, ZHL-42W+)  and delivered to the NV centre using a gold wire (diameter, $25~\mu$m) positioned accross the sample.

The single NV spin we employ for magnetometry is embedded at the apex of a nanopillar on a diamond cantilever (Fig.\,1d of main text). Such an all diamond scanning probe tip is obtained in a series of fabrication steps, including low energy ion implantation, electron beam lithography and reactive ion etching\,\cite{Maletinsky2012}. In order to ensure optimal sensitivity and nanoscale resolution for magnetic imaging, NV centres are implanted at energies of $6~$keV corresponding to an implantation depth of $10\pm8~$nm (according to SRIM simulations\,\cite{Ziegler2010}), in good agreement with the NV-to-sample standoff distance we determine from our data.

\section{S2. Sample growth}
\label{SectSample}

We fabricated epitaxially grown $c$-axis oriented YBa$_2$Cu$_3$O$_{7-\delta}$ (YBCO) thin films on SrTiO$_3$ (STO) single crystal (001)-oriented substrates by pulsed laser deposition (PLD), followed either by in-situ electron-beam-evaporation of Au at room temperature (sample A) or by in-situ epitaxial growth of an STO cap layer by PLD (sample B).
For details on PLD growth of our YBCO films on STO substrates, and their structural and electric transport properties see Refs.~[\onlinecite{Werner10,Scharinger12}].
In brief, our YBCO films typically yield $0.1^\circ$ full width half maximum of the rocking curve at the (005) x-ray diffraction peak, have an inductively measured transition temperature $T_c=89\,$K and normal state resistivity $\rho\approx 50\,\mu\Omega$cm at $T=100\,$K.

The thicknesses 
of the samples were determined through in-situ high-pressure reflection high-energy electron diffraction (RHEED) 
for the YBCO and STO films. With the $c$-axis lattice parameters, as measured via x-ray diffraction, we then obtain the thicknesses for the YBCO and STO films quoted in the text. The Au growth rate was determined via a quartz-crystal monitor. 
For all layers, we estimated the error in our thickness determination to $2-3\,\%$.
%

\section{S3. Magnetic flux in the isolated vortex image of Fig.3a in the main text}
\label{SectFlux}

In this section, we detail how we determined the experimentally measured flux in image Fig.3a of the main text and how we compared the resulting value to our expectations from a Pearl vortex.

Our experiments for Fig.3a of the main text were performed in an external offset bias field $B_{\rm offset}\approx0.1~$mT, which we applied to the sample in the $y$-direction. This bias field was necessary to perform a sign-sensitive measurement of $\vec{B}_{\rm vortex}$ --- without such a bias field, positive and negative field projections onto the NV axis could not be distinguished by virtue of time-reversal symmetry.
The value of $B_{\rm offset}$ was then subtracted from the raw data to yield the data presented in the main text.
Due to potential, small uncertainties in the calibration of our magnetic field coils, we extracted the exact values of $B_{\rm offset}$ from our fit to the date and the Pearl model.
In these fits, we set the flux quantum $\Phi_0=2.07~$mT$\mu$m$^2$ as a fixed quantity and varied $B_{\rm offset}$, Pearl length $\Lambda$ and the lateral and vertical position of the vortex with respect to the NV ($y_{\rm vortex}$ and $h_{NV}$)  as fit parameters. The value for $B_{\rm offset}$ obtained from this procedure was subtracted from the raw data to yield the data-sets for $\vec{e}_{\rm NV}\cdot\vec{B}_{\rm vortex}$ shown in Fig.3 and Fig.4 of the main text.

\begin{figure}[t]
\includegraphics[width = 8.6 cm]{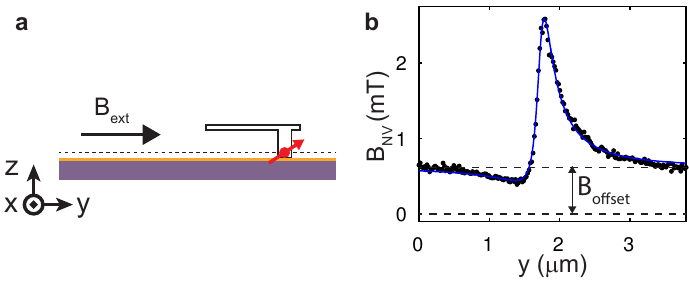}
\caption{The external B-field $B_{\rm ext}$ (a) causes a $B_{\rm offset}$ in the measured field (b), allowing to detect negativ B-field projections. }
\label{fig_Boffset}
\end{figure}

\begin{figure}[t]
\includegraphics[width = 6.5 cm]{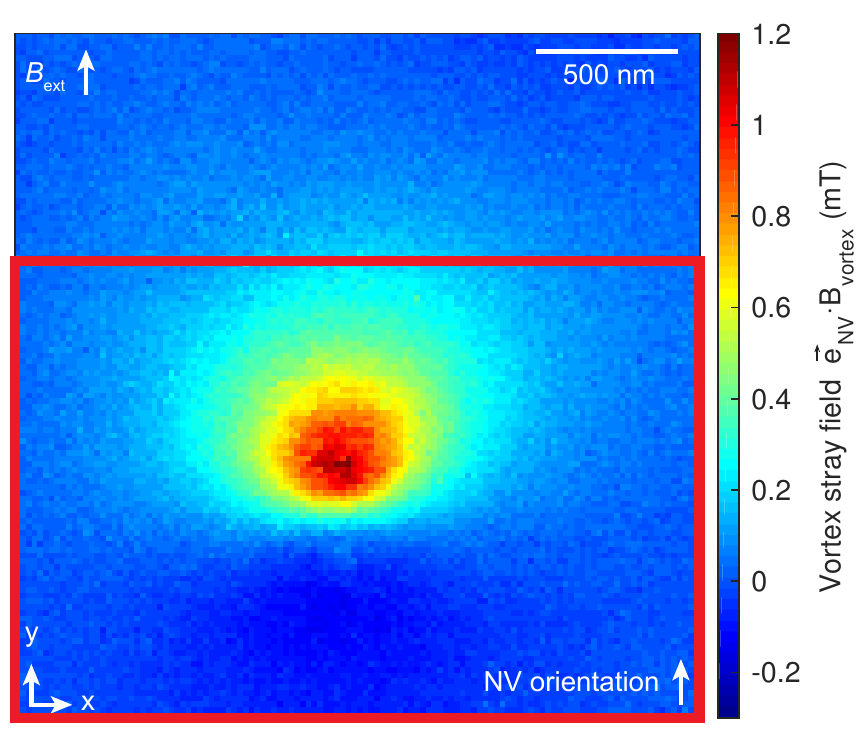}
\caption{Reproduction of Fig.3a of the main paper with the relevant integration-range for flux calculation indicated in red. We determined a magnetic flux of  0.79 $\pm$0.1 mT$\mu$m$^2$ through the red rectangle, which lies within $5\%$ of the expected, calculated flux (0.83 mT$\mu$m$^2$).}
\label{fig_flux}
\end{figure}

The experimentally measured magnetic flux $\Phi_{NV}$ was extracted from our data by numerical integration. For integration over an infinite scan plane, one would epxect
%

\begin{eqnarray}
\label{EqFlux}
\Phi_{NV} 	&=& \iint \vec{e}_{\rm NV} \cdot \vec{B}\;\mathrm{d}x \mathrm{d}y \\
		&=& \vec{e}_{\rm NV} \cdot \iint \vec{B}\;\mathrm{d}x \mathrm{d}y \\
		&=& e_{{\rm NV},z} \iint B_z\;\mathrm{d}x \mathrm{d}y \\
		&=& e_{{\rm NV},z} \Phi_0 = \frac{1}{\sqrt{3}} \Phi_0
\end{eqnarray}

where $\vec{e}_{\rm NV} = \left(0, \sqrt{\frac{2}{3}},\frac{1}{\sqrt{3}}\right) $ is the unit-vector along the NV axis. Importantly, the $x$ and $y$ components of the integral $ \iint B\;\mathrm{d}x \mathrm{d}y$ both yield zero, since $B_x$ and $B_y$ are odd functions in $x$ and $y$.
Importantly, in integrating our experimental data, we need to choose an integration area, which is centred around the vortex core (whose position we determine from the fit), in order for these $x$ and $y$ components to cancel out.  The integration range we chose to that end is indicated in red in Fig.\,\ref{fig_flux} .
The flux we determined by integrating $\vec{e}_{\rm NV}\cdot\vec{B}_{\rm vortex}$ over this area amounts to $\Phi_{0, {\rm meas}}=\sqrt{3}\Phi_{NV, {\rm meas}}=0.79\pm0.1~$mT$\mu$m$^2$, the value stated in the main text. The stated uncertainty results from uncertainties in determining $B_{\rm offset}$ from our fit.

For comparison, we numerically evaluated the integral in eq.(\ref{EqFlux}) for a Pearl vortex over the $1.6~\mu$m$\times 2.4~\mu$m integration range as used for the data for a Pearl vortex described by our initial fitting parameters (i.e. $h_{\rm NV}=162~$nm and Pearl length $\Lambda=840~$nm). This procedure yielded an expected flux $0.83~$mT$\mu$m$^2$, within $5\%$ of the experimentally measured value. We note that this agreement is surprisingly good: Nonlinearities or mis-calibrations of our piezo scanners enter to second order into the measured flux, i.e. a $5\%$ error in the flux could be caused by a $2\%$ error in the calibration of our piezo scanners.

\section{S4. NV orientation and magnetic field alignment}
\label{SectAlignment}

\begin{figure}[t]
\includegraphics[width = 8.6 cm]{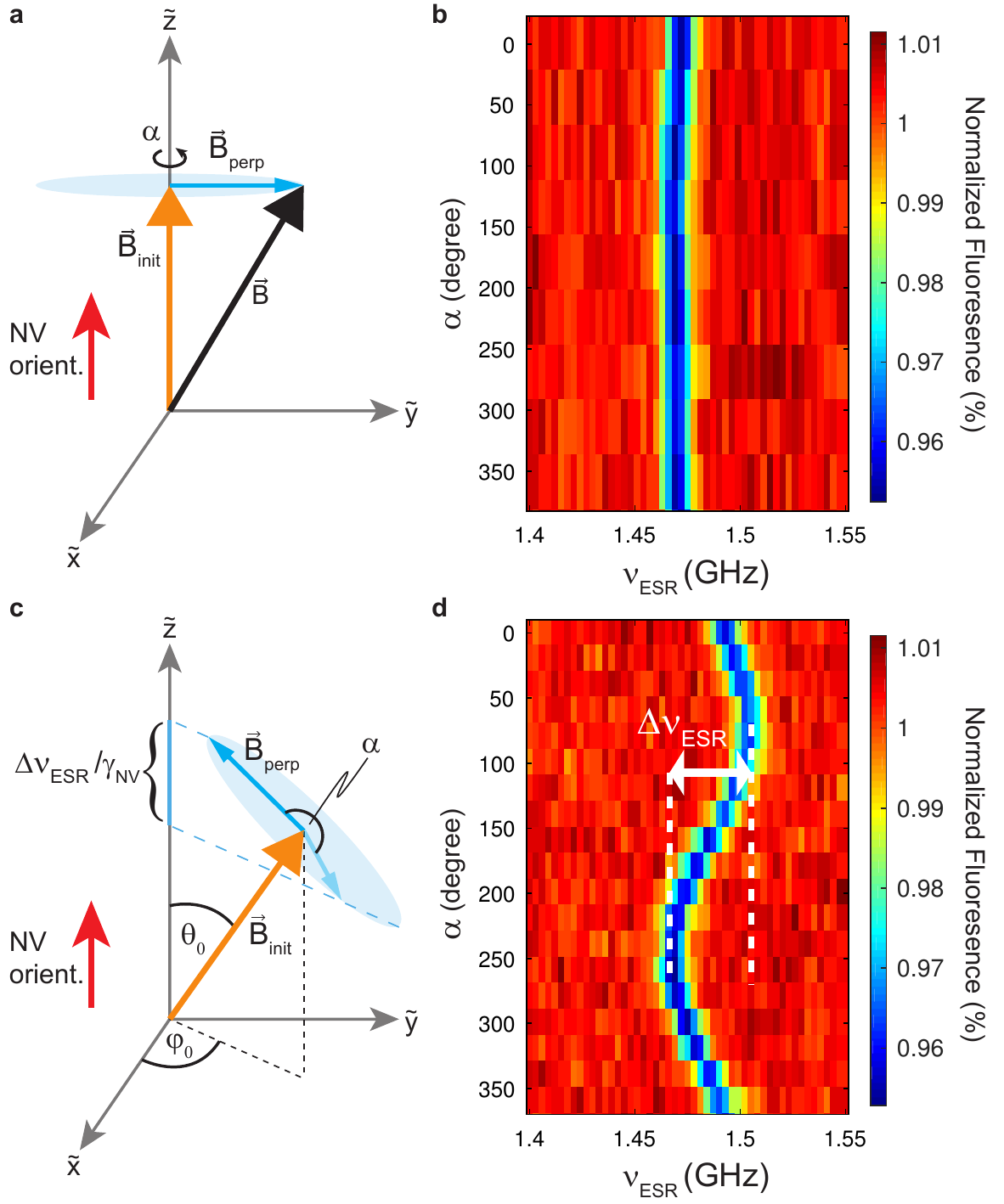}
\caption{(a) Sketch of magnetic field configurations for the case where $\vec{B}_{\text{init}}$ ($\vec{B}_{\text{perp}}$) is parallel (perpendicular) to the NV axis. (b) NV ESR as a function of azimuthal angle $\alpha$ for the situation depicted in a. For this case, where $\vec{B}_{\text{init}}$ is aligned with the NV axis, no variations of ESR frequency with $\alpha$ are observed. (c) Same as in a, but with $\vec{B}_{\text{init}}$ and $\vec{B}_{\text{perp}}$ at oblique angles with respect to the NV axis. (d) NV ESR as a function of azimuthal angle $\alpha$ for the situation depicted in c. Here, the NV ESR frequency shows strong dependance on the rotation angle $\alpha$. The observed dependance allows for the determination of the NV axis (see text).}
\label{figS4}
\end{figure}

In order to correctly interpret the magnetic field images obtained in the main text with the NV, precise knowledge of the orientation of the NV spin quantisation axis with respect to the sample coordinates is important.
We therefore developed a method based on our $3-$dimensional magnetic field control to determine this orientation in situ, without any prior assumptions.
The experimental method we employed to that end will be described in the following and is based on the fact that only magnetic fields applied {\em parallel} to the NV quantisation axis $\vec{e}_{\rm NV}$ couple to the NV spin (while fields applied {\em perpendicular} to $\vec{e}_{\rm NV}$ only lead to a weak, second.order shift of the NV spin energies).

Our method relies on controlled rotations of external magnetic fields in space and the effect such rotations have on the NV's ESR frequency. 
If a magnetic field $\vec{B}_{\text{init}}$ is applied parallel the NV axis (which for simplicity, we here set to the vertical axis, $\tilde{z}$, Fig.\,\ref{figS4}a), it will shift the ESR resonance frequency by $\Delta\nu_{\text{ESR}} = \gamma_{\text{NV}} |\vec{B}_{\text{init}}|$ with $\gamma_{\text{NV}}=28~$MHz/mT. If now a  
weak, orthogonal field $\vec{B}_{\text{perp}}$ is additionally applied, $\vec{B}_{\text{perp}}$ will not affect $\Delta\nu_{\text{ESR}}$, irrespective of the orientation of $\vec{B}_{\text{perp}}$ (parametrised by an angle $\alpha$, c.f. Fig.\,\ref{figS4}a) in the plane orthogonal to $\vec{B}_{\text{init}}$. 
The insensitivity of $\Delta\nu_{\text{ESR}}$ on $\alpha$ (as shown in Fig.\,\ref{figS4}b) is therefore a good indicator for whether $\vec{B}_{\text{init}}$ is aligned with the NV axis.
We employ a two-step iterative procedure that allows us to test this criterion and if it is not fulfilled, determine a correction to $\vec{B}_{\text{init}}$ that aligns the field closer to the NV axis.


For our alignment procedure, we start with a first approximate guess (based on crystalline directions of our diamond cantilever) for the NV direction and set $\vec{B}_{\text{init}}$ close to this direction (Fig.\,\ref{figS4}c). We then write $\vec{B}_{\text{init}}$ as 
\begin{equation}
\vec{B}_{\text{init}} = B_{\text{init}} \begin{pmatrix} \sin(\theta_{\text{0}}) \cos(\varphi_{\text{0}}) \\ \sin(\theta_{\text{0}}) \sin(\varphi_{\text{0}}) \\ \cos(\theta_{\text{0}}) \end{pmatrix}
\end{equation} 
where $B_{\text{init}}$, $\theta_{\text{0}}$ and $\varphi_{\text{0}}$ are spherical coordinates. We then apply a field $\vec{B}_{\text{perp}}$ perpendicular to $\vec{B}_{\text{init}}$ with amplitude $B_{\text{perp}}$. Rotating $\vec{B}_{\text{perp}}$ around $\vec{B}_{\text{init}}$ by an angle $\alpha$ yields a total field $\vec{B}(\alpha)$
\begin{equation}
\vec{B}(\alpha) = B_{\text{init}}\ \vec{e}_{\text{i}} +  \mathcal{R}_{\vec{B}_{\text{init}}}(\alpha) \cdot B_{\text{perp}}\ \vec{e}_{\text{p}}
\end{equation}
where $\mathcal{R}_{\vec{B}_{\text{init}}}(\alpha)$ is the rotation matrix by angle $\alpha$ around $\vec{e}_{\text{i}}$, and $\vec{e}_{\text{i(p)}}$ are the unit vectors along $\vec{B}_{\text{init}}$ ($\vec{B}_{\text{perp}}$).

Only the component of $\vec{B}(\alpha)$ which is parallel to the NV quantization axis $\tilde{z}$ affects $\Delta\nu_{\text{ESR}}$, which is therefore given as
\begin{eqnarray*}
\Delta\nu_{\text{ESR}}(\alpha) 	&=& \gamma_{\text{NV}}\ \vec{e}_{\text{NV}}\cdot \vec{B}(\alpha) \\
						&=& \gamma_{\text{NV}}\ \left[B_{\text{init}}\cos(\theta_{\text{0}}) - B_{\text{perp}}\sin(\theta_{\text{0}})\ \sin(\alpha)\right]
\end{eqnarray*}
The sinusoidal behavior of $\Delta\nu_{\text{ESR}}$ on $\alpha)$ for a slightly misaligned field $\vec{B}_{\text{init}}$ can be seen directly in Fig.\,\ref{figS4}d.

\begin{figure}[t]
\includegraphics[width = 5 cm]{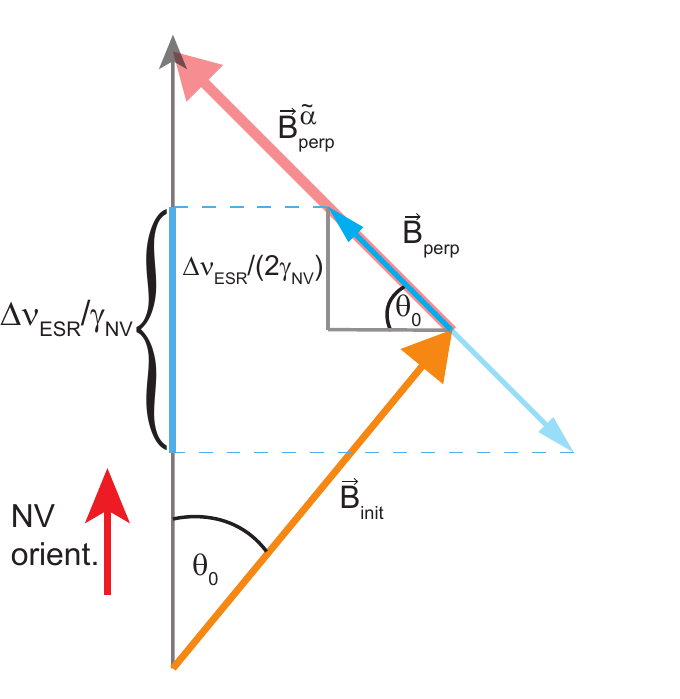}
\caption{Detailed, two-dimensional view of the schematic presented in Fig.\,\ref{figS4}c. The figure illustrates how misalignment angle $\theta_0$ and perpendicular field strength $\left|B_{perp}^{\tilde{\alpha}}\right|$ can be determined from the dataset in Fig.\,\ref{figS4}d, in order to compensate the initial misalignment in magnetic field.}
\label{figS5}
\end{figure}

Analysing the data-set in Fig.\,\ref{figS4}d allows us to align the magnetic field to the NV quantization axis. Fig.\,\ref{figS5} shows that a perpendicular field $B_{\rm perp}^{\tilde{\alpha}}$ added with appropriate angle $\tilde{\alpha}$ and amplitude $\left|B_{perp}^{\tilde{\alpha}}\right|$ to  $\vec{B}_{\text{init}}$ yields the desired field which is aligned with the NV centre. The angle $\tilde{\alpha}$ is readily identified by the criterion that it is the angle that maximizes $\Delta\nu_{\text{ESR}}$, which for Fig.\,\ref{figS4}d yields $\tilde{\alpha}\approx240^{\circ}$.

Using basic trigonometric relations one then finds from Fig.\,\ref{figS5}
\begin{equation}
\left|B_{perp}^{\tilde{\alpha}}\right| =  B_{\text{init}} \cdot \tan(\theta_{\text{0}})
\end{equation} 
where $\theta_{\text{0}}$ is the initial misalignment angle
\begin{equation}
\theta_{\text{0}} = \arcsin\left(\frac{\Delta\nu_{\text{ESR}}}{2\ \gamma_{\text{NV}}\ B_{\text{perp}}}\right)
\end{equation} 
with $\Delta\nu_{\text{ESR}}$ is the maximum splitting measured in Fig.\,\ref{figS4}d

\section{S5. AFM performance}
\label{SectAFM}

In order to verify proper tip-to-sample distance control and determine typical vibration levels of our cryostat, we performed atomic force microscope (AFM) imaging during each of our scans. We here present a typical AFM image taken on sample A during acquisition of the data shown in Fig.\,2b of the main text. The image shows a slight corrugation pattern typical for the samples we investigated. From the image, we determine an AFM roughness of the sample of $1.5~$nm and vibration levels much below that value within the acquisition bandwidth. The commercial AFM system was specified at $10$pm/$\sqrt{\rm Hz}$, corresponding to an RMS amplitude of $100~$pm for a sampling time of $5~$ms.

\begin{figure}[t]
\includegraphics[height = 4.25 cm]{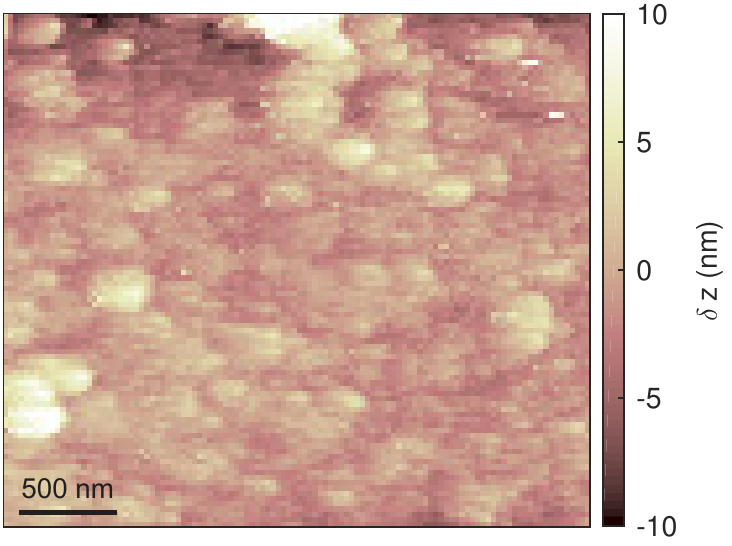}
\caption{AFM scan of sample A performed during acquisition of data shown in Fig.\,2b of the main text. From the image, we determine an AFM roughness of the sample of $1.5~$nm.}
\label{fig_SOM_AFM}
\end{figure}

\section{S6. Invasiveness of NV magnetometry}
\label{SectInvasiveness}

\begin{figure}[h!]
\includegraphics[width = 8.6 cm]{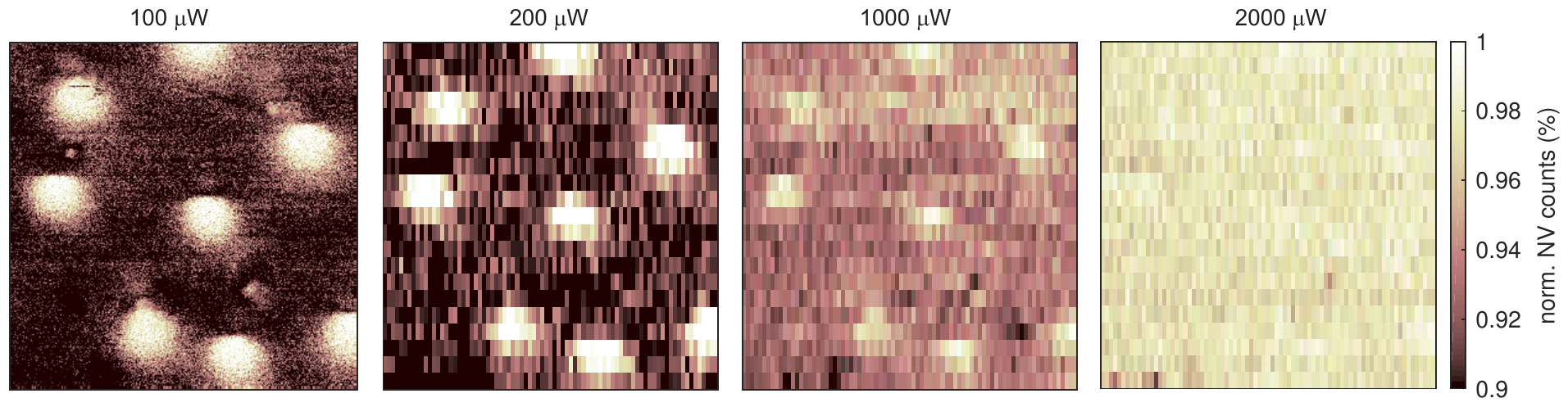}
\caption{Iso-magnetic field image taken at increased laser powers (see labels) on an ensemble of vortices in sample A. The vortices were imaged at $B=0$ after field-cooling in $B^{\rm f.c.}_{\rm z}=0.4~$mT. No "vortex dragging" was observed up to $2~$mW. Imaging contrast deteriorates for increasing laser power and eventually vanishes at $2~$mW due to the reduction of ESR contrast with excitation power\,\cite{Dreau2011}.}
\label{fig_SOM_highpower}
\end{figure}

As stated in the main text, we performed iso magnetic field imaging of vortices for varying green laser excitation powers to test the invasiveness of our method. If the excitation laser would lead to significant heating, we would expect the vortices to be "dragged" to different positions by the scanning probe. We did not observe any such vortex dragging in our experiments up to the highest laser powers employed ($2~$mW). Fig.\,\ref{fig_SOM_highpower} shows a sequence of vortex images conducted at the same location of the superconductor for laser powers of $100~\mu$W, $200~\mu$W, $1000~\mu$W and $2000~\mu$W. For increasing laser powers, the imaging contrast deteriorates due to the decreasing NV ESR contrast at these high power-levels\,\cite{Dreau2011}. The last image taken at $2000~\mu$W hardly shows any imaging contrast, but we verified after the scan that vortices did not move from their original location.

\end{document}